\documentclass[useAMS,referee,usenatbib]{biom}
\usepackage{amsmath, euscript}
\usepackage{amssymb} 
\usepackage{epsfig}  
\usepackage{multirow} 
\usepackage{multicol} 
\usepackage{float}
\usepackage{booktabs}
\usepackage[compact]{titlesec}

\newcommand\blfootnote[1]{%
  \begingroup
  \renewcommand\thefootnote{}\footnote{#1}%
  \addtocounter{footnote}{-1}%
  \endgroup
}

\titlespacing{\section}{0pt}{*0}{*0}
\titlespacing{\subsection}{0pt}{*0}{*0}
\usepackage{hyperref}
\def\bSig\mathbf{\Sigma}

\newcommand{\BF}{\boldsymbol}

\usepackage[figuresright]{rotating}
\usepackage{color}

\def\Hat{\widehat}

\def\Bar{\overline}

\title[Bayesian Functional Data Analysis]{Bayesian Functional Data Analysis over Dependent Regions and Its Application for Identification of Differentially Methylated Regions}

\author
{Suvo Chatterjee\emailx{suvo.chatterjee@nih.gov} \\
Epidemiology Branch, Division of Intramural Population Health Research,\\
\textit{Eunice Kennedy Shriver} National Institute of Child Health and Human Development, \\
National Institutes of Health, Bethesda, MD 20892, USA. 
\and 
Shrabanti Chowdhury\emailx{shrabanti.chowdhury@mssm.edu} \\
Department of Genetics and Genomic Sciences and Icahn Institute for Data Science and Genomic Technology, \\
Icahn School of Medicine at Mount Sinai, New York, NY, 10029, USA.
\and
Duchwan Ryu$^{\dagger}$\emailx{dryu@niu.edu}\\
Department of Statistics and Actuarial Science,\\
Northern Illinois University, Illinois, U.S.A.
\and
Sanjib Basu\emailx{sbasu@uic.edu} \\
Department of Epidemiology and Biostatistics,\\
University of Illinois at Chicago, Illinois, U.S.A.
}

\begin{document}

\label{firstpage}

\begin{abstract}
We consider a Bayesian functional data analysis for observations measured as extremely long sequences. Splitting the sequence into a number of small windows with manageable length, the windows may not be independent especially when they are neighboring to each other. 
We propose to utilize Bayesian smoothing splines to estimate individual functional patterns within each window and to establish transition models for parameters involved in each window to address the dependent structure between windows.
The functional difference of groups of individuals at each window can be evaluated by Bayes Factor based on Markov Chain Monte Carlo samples in the analysis.
In this paper, we examine the proposed method through simulation studies and apply it to identify differentially methylated genetic regions in TCGA lung adenocarcinoma data.

\blfootnote{$^{\dagger}$Corresponding author}

\end{abstract}
%

\begin{keywords}
Bayesian smoothing splines, Dynamic weighted particle filter, Differentially methylated region
\end{keywords}


\maketitle

\section{Introduction}
\label{s:intro}

DNA methylation is an epigenetic modification in the human DNA when a methyl group gets attached to the cytosine of CG dinucleotide, that is called CpG site, forming a methyl cytosine. Ideally these locations are free from any methylation which in turn allows the normal process of gene transcription, expression and regulation \citep{2}. Under CpG methylation this usual process is hindered causing disruption to the regular cell functioning \citep{3,4}.
DNA hypomethylation or hypermethylation cause the activation of oncogenes that cause tumor, or the silencing of tumor suppressor genes, respectively. Several researches have shown the association of differential DNA methylation patterns to various cancers and also to other chronic diseases \citep{1, 5, 6, 7, 8}.

The detection of differentially methylated regions (DMRs) has become fairly important in the last decade for locating the genes in such DMRs. One approach is to identify single sites that show differential methylation. Although methods for site-wise detection are more developed, analyses based on genomic regions may lead to substantially improved results \citep{7}. Examining regions known to impact transcriptional regulation, such as promoters and enhancers, would further improve the likelihood of regions to be identified as being differentially methylated which mediate a biological pathway or function \citep{7}.
The primary objective of several microarray based studies is to understand how the structural pattern of association \citep{60} among certain variables changes over the genomic locations. 
%
The literature on DMR identification methods includes \cite{22, 23, 24, 25, 26, 27, 28}, among others.

With the advancement in microarray technology, methylation rates can be observed in much higher resolutions, e.g., over 450K CpG locations.
Although being highly popular in methylation studies the packages summarized in \cite{37} come with several disadvantages. Most of them use traditional multivariate techniques to detect DMRs that are incapable of handling the high dimensionality, high measurement error, missing values and high degree of correlation in methylation rates among neighboring CpG sites \citep{eckhardt11}, which are some of the inherent components in  methylation data.
%
%
Hence, the traditional multivariate methods may result in spurious statistical analysis and low powered statistical tests by disregarding smooth functional behavior of the underlying data generating process.

The functional data analysis (FDA) proposed in \cite{10}, views these measurements as realizations of continuous smooth dynamic function evolving over space or time. In particular, the FDA fits these multivariate methylation data to smooth curves represented as linear combinations of suitably chosen basis functions \citep{11}. Such smooth curve enables imputation of missing values, helps in removal of the high noise, models data with irregular time sampling schedules, and accounts for any significant correlation among the observations \citep{Ryu16}. The Bayesian FDA fits a flexible Bayesian nonparametric regression model to the sequence of methylation rates and conducts statistical inference based on the fitted curves within a Bayesian framework.

To identify DMRs addressing the issues of large dimensionality, missingness and correlation of methylation rates from neighboring CpG sites, we consider genomic windows containing CpG sites by segmenting whole genome according to a certain base pairs (bp) distance between neighboring CpG sites. By examining the differential methylation profiles among two groups found in these genomic windows we identify DMRs.
It should be noted that the methylation rates from a window may associate with those from neighboring windows since the methylation rates on individual CpG sites can be dependent on each other, as \cite{eckhardt11} pointed out.

There are several methods proposed to model the dependency among the parameters in a dynamic process by transition models \citep{60} that include the auto-correlation, cross-lagged structural models \citep{40}, the auto-regressive moving averages, the cross spectral or coherence analysis \citep{54}, the dynamic factor analysis \citep{56} and the structural equation models. 
We consider the transition models for parameters that control the functional pattern of methylation rates at each window, and utilize the dynamically weighted particle filter \citep{Liang2002,Ryu2013} for efficient computation. 
Based on populations of parameters from the particle filter, we examine the Bayes factor to determine if the window is differentially methylated.

%

We  performed simulation  studies  to  benchmark  the  performance  of  our  method  with  the  popularly  used  existing bump-hunting method \citep{jaffe}. Also as an application of the proposed model, we perform DMR identification analysis over the whole genome on TCGA lung Adenocarcinoma data. The results from both simulation studies and real data analysis showed that our proposed approach is effective in finding the true DMRs while effectively controlling for the number of false positives.

The rest of the article is structured as follows. Section 2 introduces the proposed novel methods to model methylation values. Section 3 shows a comparison of the simulated data for the proposed methods with bumphunter. Section 4 shows the data analysis and important findings upon the application of the proposed dependent method to lung Adenocarcinoma data. Finally, a discussion is provided in Section 5.

\section{Bayesian Functional Data Analysis}
\label{sec:model}

In this section, we propose a Bayesian functional data analysis (BFDA) within and between windows and an identification of differentially methylated region (DMR) by utilizing Bayes factor at each window.

\subsection{BFDA within a window}
\label{sec:withinBFDA}
%
We configure the windows by splitting a long sequence of data collected from each individual. 
%
%
Regarding the DNA methylation data measured on CpG sites over whole human genome, the genomic window of CpG sites can be determined by the number of CpG sites in the window, the gap of genomic coordinates of neighboring CpG sites or other reasonable rules. 

We consider the methylation data measured at a window with $n$ CpG sites from $m_k$ individuals in the group $k$, $k=1,\ldots,G$. 
%
Let $Y_{ijk}$ denote the log-transformed methylation rate, referred as M-value in \cite{25}, at the CpG site $i$ from the individual $j$ in the group $k$ as 
\begin{eqnarray}\label{transform}
Y_{ijk} =\log\left(\frac{\beta_{ijk}+c}{1-\beta_{ijk}+c}\right),~~~i=1,\ldots,n,~j=1,\ldots,m_k,~k=1,\ldots,G,
\end{eqnarray}
where $\beta_{ijk}$ is the methylation rate and $c$ is an offset value.  
%

We utilize a Bayesian nonparametric regression in the BFDA for the sequence of measured methylation rates over CpG sites, $Y_{1jk},\ldots,Y_{njk}$, from the individual $j$ in group $k$. The typical features of methyulation rates include high variability, nonperiodic behavior, correlation and complex patterns over CpG sites.
Without assumption of specific functional form, denoting the mean functional value of methylation rate as $g_k(x_i)$ at the CpG site $i$ over the individuals in group $k$, we can model the transformed methylation rates by the following regression model:
\begin{equation}
    \label{Equation11}
    Y_{ijk}=g_k(x_i)+\delta_{ik} + \epsilon_{ijk},~
    i=1,\ldots,n,~
    j=1,\ldots,m_k,~
    ~k=1,\ldots,G,
\end{equation}
where $\delta_{ik}$ is the random component induced by the group $k$, or the discrepancy of $g_k(\cdot)$, and $\epsilon_{ijk}$ is the random error from the individual $j$ in the group $k$, and $\delta_{ik}$ and $\epsilon_{ijk}$ are mutually independent with zero means and constant variances, $\sigma_k^2$ and $\sigma_{jk}^2$, respectively.
In this paper, to investigate the functional pattern of methylation rates over CpG sites, we use the order of CpG site in the window, that is $x_i=i$, as the predictor in the model instead of its genomic coordinate.

Using the natural cubic smoothing splines that is conventional, the mean function can be described by $g_k(x)=\sum_{t=1}^Ta_{kt}B_t(x-\mu_t)$, for the natural cubic basis functions $B_t(\cdot)$ and unique knot points $\mu_t$, $t=1,\ldots,T$, where $a_{kt}$ are the coefficients of basis functions. 
The smoothing splines allow only one response at a unique design point. Using the group mean as the response, $\Bar Y_{ik}=\frac1{m_k}\sum_{j=1}^{m_k}Y_{ijk}$, for $i=1,\ldots,n$ and $k=1,\ldots,G$, the fitted curve of the smoothing splines $g_k(\cdot)$ can be found by minimizing the following penalized residual sum of square
\begin{equation}
    \sum_{i=1}^n \left\{\Bar Y_{ik}-g_k(x_{i})\right\}^2 + \alpha_k\int_{u\in \mathcal{R}} g_k^{\prime\prime}(u)^2 du,
\end{equation}
where $\alpha_k$ is a given positive smoothing penalty, $g_k^{\prime\prime}(\cdot)$ is the second order derivative of $g_k(\cdot)$ and $\mathcal{R}$ is the range of design points. Denoting the vector of functional value of the smoothing splines as $\BF g_k=[g_k(x_1),\ldots,g_k(x_n)]^T$, the penalty term can be expressed by $\int_{u\in \mathcal{R}}g_{k}^{\prime\prime}(u)^2du = \BF g_k^T\BF K\BF g_k$, where $\BF K$ is an $n\times n$ dimensional matrix with rank $n-2$. Refer to \cite{17} for details of construction of $\BF K$. 
Here, it should be noted that all individuals share the same design points that are the CpG sites and hence the group mean functions share the same $\BF K$.

As mentioned in \cite{20, Ryu11,42}, for a Bayesian approach, we take all design points as knot points and assign a singular normal prior on $\BF g_k$ that has the probability density function proportional to $\left(\frac{\alpha_k}{\sigma_k^2}\right)^{(n-2)/2}\!\!\exp\left\{-\frac{\alpha_k}{\sigma_k^2} {\BF g_k}^{T}{\boldsymbol{K}}{\BF g_k}\right\}$, 
where $\alpha_k$ is a smoothing penalty and $\sigma_k^2$ is the variance of the discrepancy of the mean function. Without loss of generality, we use $\tau_k=\frac{\alpha_k}{\sigma_k^2}$ and assign a conjugate Gamma prior, $\tau_k\sim G(A_t,B_t)$. We assign conjugate inverse Gamma priors on $\sigma_k^2$ and $\sigma_{jk}^2$, respectively, $\sigma_k^2\sim IG(A_s,B_s)$ and $\sigma_{jk}^2\sim IG(A_{s}^*,B_{s}^*)$. 
%
%
Denoting $\BF y_{jk}=(y_{1jk},\ldots,y_{njk})^T$ and $\Bar{\BF y}_k=(\Bar Y_{1k},\ldots,\Bar Y_{nk})^T$, the full conditional distributions of the parameters are given by
\begin{eqnarray*}
    \BF g_k\vert\cdot 
    &\sim& 
    N\left[(\BF I+\alpha_k\BF K)^{-1}\Bar{\BF y}_k,\,(\BF I+\alpha_k\BF K)^{-1}\sigma_k^2\right],~~~k=1,\ldots,G\\
    \tau_k\vert\cdot
    &\sim&
    G\left[\frac{n-2}2+A_t,\,\frac12\BF g_k^T\BF K\BF g_k\right],\\
    \sigma_k^2\vert\cdot
    &\sim&
    IG\left[\frac n2+A_s,\,\frac12(\Bar{\BF y}_k-\BF g_k)^T(\Bar{\BF y}_k-\BF g_k)+B_s\right],\\
    \sigma_{jk}^2\vert\cdot
    &\sim&
    IG\left[\frac n2+A_s^*,\,\frac12(\BF y_{jk}-\Bar{\BF y}_k)^T(\BF y_{jk}-\Bar{\BF y}_k)+B_s^*\right],~~~j=1,\ldots,m_k,
\end{eqnarray*}
where $\BF I$ is the $n\times n$ identity matrix. 
Using Gibbs sampler with $N$ iterations after $B$ iterations as burning time, we generate MCMC samples of $\BF g_k$ and other nuisance parameters.

\subsection{BFDA between windows over whole genome}
\label{sec:betweenBFDA}

When the windows of CpG sites are independent we may apply the BFDA discussed in the previous subsection to each window. However, the windows can be associated with each other especially when they are adjacent. In this subsection, we model the dependent structure of consecutive windows by using parameter transition model and propose to utilize dynamically weighted particle filter (DWPF) for efficient computation as in \cite{Ryu2013}.

Regarding two adjacent windows, the correlation between them may not be apparent because  there are several ways to pair a CpG site from one window with another CpG site from the other window and furthermore two windows may not have the same number of CpG sites. Instead, for the dependent windows, we consider the association of the curves that fit the responses at each window. Specifically, because the smoothing penalty and the variance of the discrepancy characterize the fitted curve in the window, we let those parameters take into account the dependent windows. Suppose there are $T$ windows and let $\tau_{t,k}$ denote the smoothing penalty and $\sigma_{t,k}^2$ denote the variance of the discrepancy,  for the window $t$, $t=1,\ldots,T$, and the group $k$, $k=1,\ldots,G$.  
Then, we may consider the following linear transition models between window $t-1$ and window $t$:

\begin{equation}\label{lintrans}
    \begin{split}
    \log(1/\tau_{t,k}) &= \log(1/\tau_{t-1,k}) + U_{t,k},\\
    \log(\sigma_{t,k}^2) &= \log(\sigma_{t-1,k}^2) + V_{t,k},
    \end{split}
\end{equation}
where $U_{t,k}$ and $V_{t,k}$ are Gaussian random errors with zero mean and constant variance, respectively.

For whole genome, regarding $T$ windows, let $\BF y_{t,k}$ denote the mean methylation rates and $\BF g_{t,k}$ denote the vector of functional values of methylation curve at all CpG sites in window $t$, $t=1,\ldots,T$, and group $k$, $k=1,\ldots,G$.
Utilizing the dynamically weighted particle filter (DWPF) for efficient computing as \cite{Ryu2013} did, we apply DWPF for parameters $\tau_{t,k}$ and $\sigma_{t,k}^2$. Let $\BF\lambda_t=(\tau_{t,1},\ldots,\tau_{t,G},\sigma_{t,1}^2,\ldots,\sigma_{t,G}^2)$ denote the vector of those parameters and $\BF y_t=(\BF y_{t,1},\ldots,\BF y_{t,G})$ and $\BF g_t=(\BF g_{t,1},\ldots,\BF g_{t,G})$ denote the vector of methylation rates and the vector of functional values over all groups, respectively.
Regarding the population of $N_t$, for $i=1,\ldots,N_t$, let $\BF\lambda_t^{(i)}$ denote the particle $i$ of $\BF\lambda_t$ and $w_t^{(i)}$ denote the weight of the particle $i$, and $\BF\Lambda_t=(\BF\lambda_t^{(1)},\ldots,\BF\lambda_t^{(N_t)})$ denote the population of all particles with weights $\BF W_{\!\!t} = (w_t^{(1)},\ldots,w_t^{(N_t)})$.

In DWPF, we use the dynamically weighted importance sampling (DWIS) algorithm with $W$-type move in the dynamic weighting step and the adaptive pruned-enriched population control scheme in the population control step. Further, we denote the lower and upper population size control bounds as $N_\mathrm{low}$ and $N_\mathrm{up}$, respectively, and the lower and upper limiting population sizes as $N_\mathrm{min}$ and $N_\mathrm{max}$, respectively. We also denote the lower and upper weight control bounds for all windows as $W_\mathrm{low}$ and $W_\mathrm{up}$, respectively.

Denoting $\BF y_{1:t}=(\BF y_1,\ldots,\BF y_t)$ and $\BF\lambda_{1:t}=(\BF\lambda_1,\ldots,\BF\lambda_t)$ and suppressing the design points that are the order of CpG sites within a window, we use the the following algorithm to collect the particles of $\BF\lambda_t$ with corresponding weights $w_t$ for window $t$, $t=1,\ldots,T$:
\begin{itemize}
    \item []\underline{window 1}
    \begin{itemize}\setlength{\itemindent}{-1em}
        \item[]Sample:
        Collect $N_1$ MCMC samples of $\BF\lambda_1$ from the posterior distribution by applying the BFDA discussed in the previous subsection after $B$ burning iterations, and set the MCMC samples at each iteration as $\Hat{\BF\lambda}_1^{(i)}$ with weight $\Hat w_1^{(i)}=1$, $i=1,\ldots,N_1=20000$. It establishes $\Hat{\BF\Lambda}_1$ and $\Hat{\BF W}_{\!\!1}$.
	    \item[]DWIS: Generate $(\BF\Lambda_1,\BF W_{\!\!1})$ from $(\Hat{\BF\Lambda}_1,\Hat{\BF W}_{\!\!1})$ 
        using DWIS, with the marginal posterior distribution $p(\BF\lambda_1\vert\BF y_1)$ as the target distribution.
    \end{itemize}
    \item []\underline{window 2}
    \begin{itemize}\setlength{\itemindent}{-1em}
	    \item[] Extrapolation: Generate $\Hat{\BF\lambda}_2^{(i)}$ from $\BF\lambda_1^{(i)}$, with the 
        extrapolation operator $q(\BF\lambda_2\vert\BF\lambda_1^{(i)}\!\!,\,\BF y_{1:2})$, and set
        $
            \Hat w_2^{(i)} = w_1^{(i)}
                \frac{p(\BF\lambda_1^{(i)}\!\!,\,\Hat{\BF\lambda}_2^{(i)}\vert\BF y_{1:2})}
                {p(\BF\lambda_1^{(i)}|\BF y_1)
                    q(\Hat{\BF\lambda}_2^{(i)}|\BF\lambda_1^{(i)}\!\!,\,\BF y_{1:2})}
        $
        for each $i=1,2,\ldots, N_1$, to establish $(\Hat{\BF\Lambda}_2,\Hat{\BF W}_{\!\!2})$.
        \item[] DWIS: Generate $(\BF\Lambda_2,\BF W_{\!\!2})$ from $(\Hat{\BF\Lambda}_2,\Hat{\BF W}_{\!\!2})$ 
        using DWIS, with the target $p(\BF\lambda_{1:2} | \BF y_{1:2})$.
    \end{itemize}
    \item[]~~~~~$\vdots$
    \item[]\underline{window $T$}
    \begin{itemize}\setlength{\itemindent}{-1em}
        \item[] Extrapolation: Generate $\Hat{\BF\lambda}_t^{(i)}$ from $\BF\lambda_{t-1}^{(i)}$, 
        with the extrapolation operator $q(\BF\lambda_t\vert\BF\lambda_{1:t-1}^{(i)},\BF y_{1:t})$ 
        and set
        $
            \Hat w_{t}^{(i)} = w_{t-1}^{(i)}
                \frac{p(\BF\lambda_{1:t-1}^{(i)},\Hat{\BF\lambda}_t^{(i)} | \BF y_{1:t})}
                {p(\BF\lambda_{1:t-1}^{(i)}\vert\BF y_{1:t-1})   
                    q(\Hat{\BF\lambda}_t^{(i)}\vert\BF\lambda_{1:t-1}^{(i)},\BF y_{1:t})}
        $
        for each $i=1,2,\ldots, N_{t-1}$, to establish $(\Hat{\BF\Lambda}_{t},\Hat{\BF W}_{\!\!t})$.
        \item[] DWIS: Generate $({\BF\Lambda}_t,{\BF W}_{\!\!t})$ from $(\Hat{\BF\Lambda}_t,\Hat{\BF W}_{\!\!t})$
        using DWIS, with the target  $p(\BF\lambda_{1:t} \vert \BF y_{1:t})$.
    \end{itemize}
\end{itemize}
At each window, the functional values of methylation rates $\BF g_t$ can be generated from its full conditional distribution $p(\BF g_t\vert\cdot)$, $t=1,\ldots,T$.
See \cite{Ryu2013} for details of DWPF.

\subsection{Identification of differentially methylated regions using Bayes factor}

We examine the differential methylation by groups at each window. In model \eqref{Equation11}, when different group mean functions are desirable to model the methylation rates, we may assess the window to be differentially methylated. Otherwise, the window is not differentially methylated and one group mean function will be enough to model the methylation rates in the window. We consider the following two models $M_1$ and $M_2$:
\begin{equation*}
    \begin{aligned}
    M_1: &~\mbox{the window has one group mean function, $G=1$ in model \eqref{Equation11},} \\
    M_2: &~\mbox{the window has more than one (say, $K$) group mean functions, $G=K$ in model \eqref{Equation11}.} 
    \end{aligned}
\end{equation*}
If $M_1$ is preferred to model the methylation rates in the window, we may conclude the window is not a differentially methylated region (DMR); whereas, if $M_2$ is preferred, we can conclude the window is a DMR.

To choose a good model for the methylation rates in the window, we utilize the posterior Bayes factor that provides more consistent results than the Bayes factor does, as \cite{Aitkin91} mentioned. Let $\BF\Theta_l$ denote all unknown quantities in model $M_l$ and $\BF y$ denote the log-transformed methylation rates in the window with the likelihood $p(\BF y\vert\BF\Theta_l)$, $l=1,2, \dots, K$, then the posterior Bayes factor to compare $M_1$ and $M_2$ can be calculated by the ratio of the marginal likelihoods as follows
\begin{eqnarray*}
    BF(M_1, M_2) 
    = 
    \frac{\int_{\BF\Theta_1} p(\BF y\vert\BF\Theta_1)p(\BF\Theta_1\vert\BF y)d\BF\Theta_1}
    {\int_{\BF\Theta_2} p(\BF y\vert\BF\Theta_2)p(\BF\Theta_2\vert\BF y)d\BF\Theta_2},
\end{eqnarray*}
where $p(\BF\Theta_l\vert\BF y)$, $l=1,2, \dots, K$, indicates the posterior density. Under the model \eqref{Equation11}, apparently $p(\BF y\vert\BF\Theta_l)$ is given by a product of Gaussian densities.
Utilizing the particles and weights for $p(\BF\Theta_l\vert\BF y)$ discussed in the previous subsection we obtain the marginal likelihoods by taking the weighted average of the likelihoods and calculate the Bayes factor for each window. To avoid the computational difficulty we use the log-scaled Bayes factor. If the Bayes factor is less than a threshold value, we prefer $M_2$ over $M_1$ and identify the window as a DMR. 

\subsection{Parameter values for simulation and real data analysis}
In this paper for our simulation studies as well as real data analysis, specifically, we consider two groups to identify differential methylation, cancer (case) group and normal (control) group, i.e., we have $G=2$. In Equation (1) we set the offset $c=0.01$. We set the hyper-parameters for the Gamma and inverse Gamma priors as $A_t=1$, $B_t=1000$ and $A_s=B_s=A_{s}^*=B_{s}^*=1$. We use Gibbs sampler with $N=20,000$ iterations and $B=1000$ iterations as burning time. In DWPF, we set the lower and upper population size control bounds as $N_\mathrm{low}=15000$ and $N_\mathrm{up}=25000$, respectively, and the lower and upper limiting population sizes as $N_\mathrm{min}=10000$ and $N_\mathrm{max}=30000$, respectively. We also set the lower and upper weight control bounds for all windows as $W_\mathrm{low}=e^{-5}$ and $W_\mathrm{up}=e^{5}$, respectively.

\section{Simulation Studies}
\label{sec:simul}

We examine the performance of the proposed functional data analysis for identification of deferentially methylated regions (DMRs) through simulation studies. We consider 25 subjects from the control group and 50 subjects from the case group and simulate methylation rates over 10 dependents windows. At each window the sequence of methylation rates from each subject is generated by a random function and autocorrelated random errors along with 100 equally spaced CpG sites. As the random functions of each group, we consider a group mean function and add random fluctuations.


For the subject at window $t$ from group $k$, we consider the following sinusoidal group mean functions, $g_{t,k}(x)$, $t=1,\ldots,10$;~$k=1,2$, 
\begin{equation*}
\begin{array}{cclccl}
	g_{1,1}(x) &\!\!=\!\!& 1/[1\!+\!\exp\{-\sin(2\pi x)\}],&
	g_{1,2}(x) &\!\!=\!\!& 1/[1\!+\!\exp\{-\sin(2\pi x)\}],\\
	g_{2,1}(x) &\!\!=\!\!& 1/[1\!+\!\exp\{-\sin(\pi x)\}],&
    g_{2,2}(x) &\!\!=\!\!& 1/[1\!+\!\exp\{-\sin(2\pi x)\}],\\
	g_{3,1}(x) &\!\!=\!\!& 1/[1\!+\!\exp\{-\sin(\pi+\pi x)\}],&
    g_{3,2}(x) &\!\!=\!\!& 1/[1\!+\!\exp\{\sin(\pi+\pi x)+1\}],\\
	g_{4,1}(x) &\!\!=\!\!& 1/[1\!+\!\exp\{-\sin(\frac{\pi x}{2})\}],&
    g_{4,2}(x) &\!\!=\!\!& 1/[1\!+\!\exp\{\sin(\frac{\pi x}{2})-1\}],\\
	g_{5,1}(x) &\!\!=\!\!& 1/[1\!+\!\exp\{-\sin(\frac{\pi+\pi x}{2})\}],&
    g_{5,2}(x) &\!\!=\!\!& 1/[1\!+\!\exp\{\frac{\sin(\frac{\pi}{2}+2\pi x)\}-1}{2}],\\
	g_{6,1}(x) &\!\!=\!\!& 1/[1\!+\!\exp\{-\sin(\pi+\frac{\pi x}{2})\}],&
    g_{6,2}(x) &\!\!=\!\!& 1/[1\!+\!\exp\{-\sin(\pi+\pi x)\}],\\
	g_{7,1}(x) &\!\!=\!\!& 1/[1\!+\!\exp\{-\sin(\frac{3\pi+\pi x}{2})\}],&
    g_{7,2}(x) &\!\!=\!\!& 1/[1\!+\!\exp\{\sin(\frac{3\pi+\pi x}{2})+1\}],\\
	g_{8,1}(x) &\!\!=\!\!& 0.5,&
    g_{8,2}(x) &\!\!=\!\!& 0.5,\\
	g_{9,1}(x) &\!\!=\!\!& 0.75,&
    g_{9,2}(x) &\!\!=\!\!& 0.25,\\
	g_{10,1}(x) &\!\!=\!\!& 0.25,&
    g_{10,2}(x) &\!\!=\!\!& 0.4,
\end{array}
\end{equation*}
where $x=x_1,\ldots,x_{100}$ and $x_i$, $i=1,\ldots,100$, are equally spaced sites from $x_1=0$ to $x_{100}=1$.

To generate individual random functions, we add independent normal random error at each site from $N(0,0.2^2)$, apply wavelet denoising with Daubechies 10 and parameter $\alpha=3$, and transform the denoised functions back to the original scale, based on the logit transformed group mean functions. 
The decomposition levels in the denoising are 5, 4, 3, 2, 3, 4, 5, 4, 3 and 2 from the window 1 to the window 10, where the larger level brings the smoother curve. 
By adding random errors from AR(1) to the individual functions with different variances such that $0.4^2$, $0.6^2$, $0.8^2$, $1$, $0.8^2$, $0.6^2$, $0.4^2$, $0.6^2$, $0.8^2$ and 1 for the window 1 through the window 10, respectively, we simulate the sequence of methylations on sites over 10 windows for each subject. It should be noted that the neighboring windows are dependent to each other by the systematic denoising of decomposition levels and the variance of errors as well as the autocorrelated errors. Figure 1 demonstrates an example of random functions and simulated data for 10 windows.

We have generated 100 sets of simulated methylation rates and compare the performance of three methods in the identifications of DMRs from the generated data. First, by assuming independent windows, we apply the proposed method denoted by Independent method. Next, by allowing dependency among neighboring windows, we examine the proposed method denoted by Dependent method. Finally, as a conventional counter part, we use the  bump-hunting approach \citep{jaffe} denoted by Bumphunter.

\begin{figure}
\begin{center}
\includegraphics[width=16cm,height=10cm,angle=0,trim=0 0 0 0,clip]{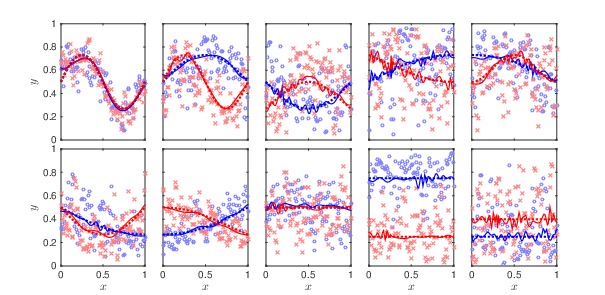}
\caption{Left to the right and first line to the second line represents window 1 to window 10. Blue color and red color indicate control group and case group, respectively. Dotted lines describe group mean functions, circles and crosses indicate the generated data based on the random functions.}  
\label{figure1}
\end{center}
\end{figure} 
 

\par In the Independent method and the Dependent method, we use a suitable threshold $c$ to identify DMRs using the log Bayes factor $\log BF(M_1,M_2)$, where $M_1$ is the model with one group mean function and $M_2$ is the model with two group mean functions.

%
%

The plots in Figure 2 shows the simulation results using 
two methods.
Figure 2 shows the distribution of the Bayes factors over 100 simulated data sets for each window under the proposed approaches for the four data sets with the respective auto-correlation 0, 0.3, 0.5 and 0.7 among the windows. In each plot the blue horizontal line represents our Bayes factor cut-off value $c$. Any window with $\log BF(M_1,M_2)$ below this cut-off is detected as a DMR. Further, in Web Appendix, supplementary figure 1 shows the distribution of the number of bumps found in each window over 100 simulations. The blue horizontal line represents the average number of bumps such that any window with at least those many bumps will be considered a DMR.

It is clear from Figure 2 that the Independent method performs the best when the CpG sites and the consecutive windows are uncorrelated and its performance deteriorates with the increase in the correlation. Table 1 describes the misclassification rates of the identification of DMRs over 100 simulations for the three methods.
As we can see the Independent method performs fairly well in detecting DMRs for windows 1, 2, 7, 9, 10 but it misclassifies windows 3, 4, 5, 6, 8 quite a lot and the missclassification rate increases with increase in the correlation among the CpG sites. 

\begin{table}
\centering
\renewcommand{\arraystretch}{2}
\caption{Misclassification rates of the independent, dependent and bumphunter methods over 100 simulations for each of the 4 data sets which were created with varying correlations of R=0, 0.3, 0.5, 0.7 among the CpG sites. The computation time depicts the average time taken to complete analysis on each window.} \label{tab:1}
\centering 
\resizebox{\textwidth}{!}{
\begin{tabular}{ccccccccccccc}
    \toprule
    \hline
    \multirow{2}{*}{\textbf{Method}} & \multirow{2}{*}{\textbf{Correlation}} & \multicolumn{10}{c}{\textbf{Window}} & \multirow{2}{*}{\textbf{Time (minutes)}}\\
  \cline{3-12}   
        & & \textbf{1} & \textbf{2} & \textbf{3} & \textbf{4} & \textbf{5} & \textbf{6} & \textbf{7} & \textbf{8} & \textbf{9} & \textbf{10} &  \\
    \midrule
    \hline
    \multirow{4}{*}{independent} & \textbf{R=0} & 0\%   & 0\%   & 0\%   & 3\%   & 20\%  & 15\%  & 0\%   & 2\%   & 0\%   & 4\%   & 112.45 \\

   & \textbf{R=0.3} & 0\%   & 0\%   & 1\%   & 6\%   & 22\%  & 22\%  & 0\%   & 25\%  & 0\%   & 5\%   & 115.12 \\
  
    & \textbf{R=0.5} & 0\%   & 0\%   & 9\%   & 14\%  & 32\%  & 24\%  & 0\%   & 39\%  & 0\%   & 21\%  & 113.58 \\
  
    & \textbf{R=0.7} & 0\%   & 8\%   & 23\%  & 38\%  & 41\%  & 28\%  & 6\%   & 56\%  & 0\%   & 23\%  & 111.33 \\
    \bottomrule
    \multirow{4}{*}{dependent} & \textbf{R=0} & 0\%   & 0\%   & 0\%   & 1\%   & 9\%   & 7\%   & 0\%   & 1\%   & 0\%   & 2\%   & 22.49 \\
    & \textbf{R=0.3} &  0\%   & 0\%   & 0\%   & 3\%   & 10\%  & 10\%  & 0\%   & 11\%  & 0\%   & 2\%   & 23.024 \\
    & \textbf{R=0.5} &  0\%   & 0\%   & 4\%   & 6\%   & 14\%  & 10\%  & 0\%   & 17\%  & 0\%   & 9\%   & 22.716 \\
    & \textbf{R=0.7} &  0\%   & 3\%   & 10\%  & 17\%  & 18\%  & 12\%  & 3\%   & 24\%  & 0\%   & 10\%  & 22.266 \\
    \bottomrule 
    \multirow{4}{*}{bumphunter} & \textbf{R=0} &  0\%   & 81\%  & 0\%   & 0\%   & 28\%  & 24\%  & 52\%  & 0\%   & 100\% & 0\%   & 7.49 \\
    & \textbf{R=0.3} & 0\%   & 89\%  & 1\%   & 0\%   & 33\%  & 36\%  & 61\%  & 0\%   & 100\% & 0\%   & 7.12 \\
    & \textbf{R=0.5} & 0\%   & 84\%  & 1\%   & 0\%   & 26\%  & 36\%  & 68\%  & 0\%   & 100\% & 0\%   & 6.99 \\
    & \textbf{R=0.7} & 0\%   & 90\%  & 16\%  & 5\%   & 28\%  & 41\%  & 78\%  & 0\%   & 100\% & 2\%   & 8.25 \\
    \hline
    \bottomrule
    \end{tabular}}
\end{table}

The misclassifications in windows 4, 5 and 6 are particularly of concern since inferring a true DMR as non-DMR leads to missing out on important genes in the neighborhood of the methylated regions. Capturing DMR shows strong association of gene with the disease.

%
As for the Dependent method, it not only outperforms the Independent method in detecting true DMRs in the presence of significant correlation among CpG sites and consecutive windows, but is also computationally at least 5 times more efficient.
This further confirms that the proposed Dependent method not only provides accurate model fittings by flexible non-parametric smooth functions accounting for the highly auto-correlated CpG sites, but also acknowledges the associations among the CpG sites from the neighboring windows through the linear transition models in a very efficient and robust way.
We also observe that the Bumphunter method performs much worse than other two methods due to its high sensitivity in sudden low variations and low sensitivity in the presence of large amount of variation and sparsity in the measurements. This property of Bumphunter method is particularly revealed for DMR detection in windows 2, 7 and 9 that have large variability in the mean methylations among normal and cancer groups. The Bumphunter method misclassifies these windows as non-DMRs almost always, over 90\% of the cases. Also, since window 8 has a very low variability in the two group means, the Bumphunter method has a 100\% accuracy in detecting this window as non-DMR.

Summarizing the simulation results, the performance of both the Independent method and the Bumphunter method drops as the correlation increases significantly among the CpG sites, whereas, the Dependent approach clearly sets a higher benchmark in consistently detecting DMRs even in the presence of strong correlation among CpG sites and the neighboring windows. 

\begin{figure}
\begin{center}
\leftskip -1cm
\includegraphics[width=18cm,height=15cm]{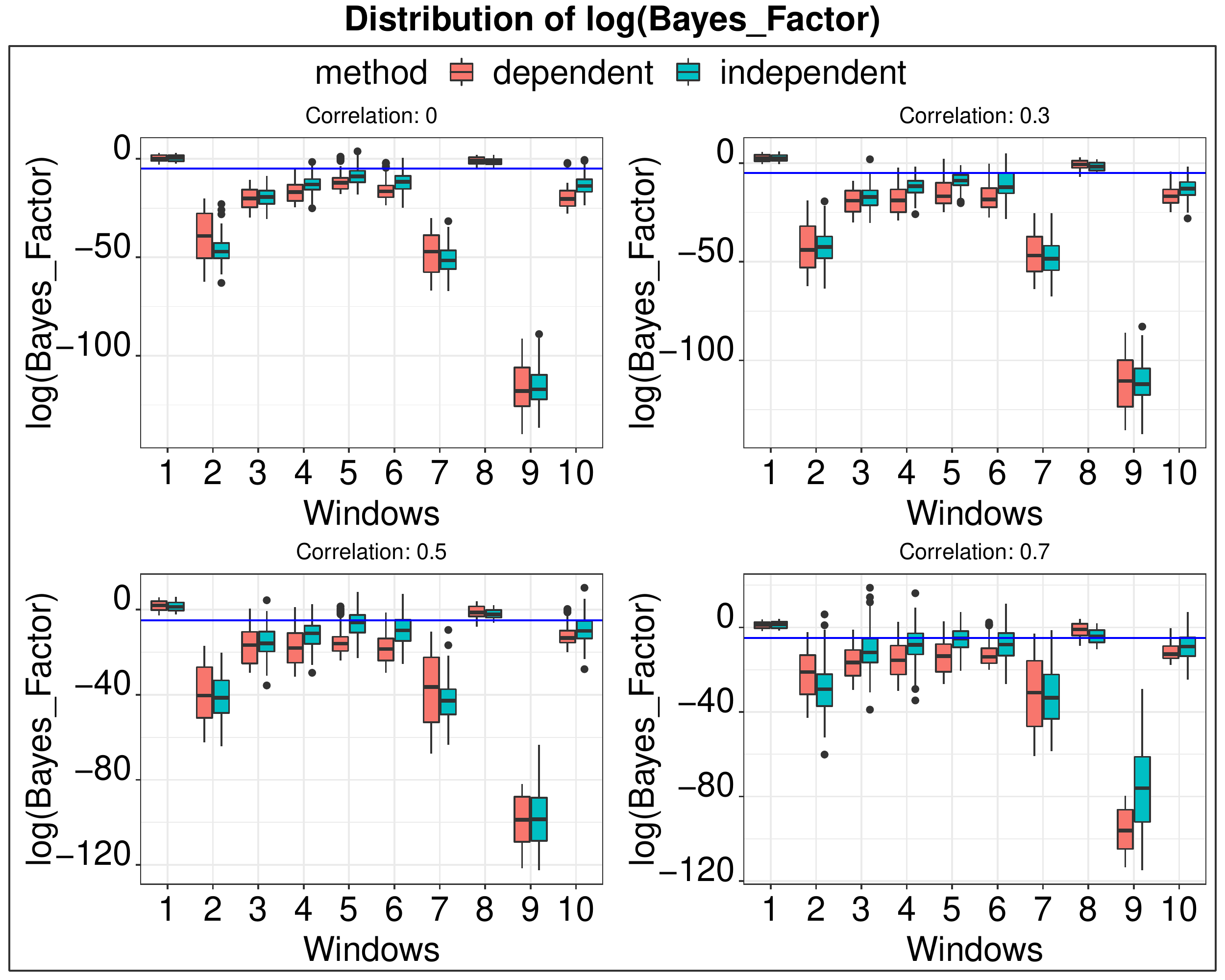}
\caption{Performance of the independent and dependent methods. The plots show the distribution of logarithm of the Bayes factors for window 1 through 10 over 100 simulations in 4 data sets, from top left to bottom right, with respective correlations $\rho$=0, 0.3, 0.5, 0.7. The horizontal blue line indicates the threshold value set at -5 and any window below this threshold is detected as a DMR.} 
\label{figure2}
\end{center}
\end{figure}

Moreover, the proposed method utilizes the functional data analysis that provides a flexible way of dimension reduction and results in more powerful detection of DMRs with the Bayes factor than the conventional multivariate approaches.
%
%
Although the Dependent method takes 3 times longer than the Bumphunter method in computation time, it is 5 times faster than the Independent method, which further shows its advantages from the scalability point of view which is often an issue encountered in Bayesian modeling.

\section{Identification of Differentially Methylated Regions for Lung Adenocarcinoma}

We illustrate our proposed approach of DMR detection using the 450K methylation data on Lung Adenocarcinoma obtained from the cancer genome atlas (TCGA) portal. This data had more than 485,512 CpG loci and encompassed 1.5\% of the total genomic CpG sites \citep{sandoval9}. There were 254 samples of cancer patients and 32 samples of normal patients with the methylation rates, denoted by $\beta-$values, which were calulated using the intensities received from the methylated and the unmethylated alleles. In the following we present the detailed analysis of DMR detection across the whole genome using our proposed approaches.

\subsection{DNA Methylation Data Analysis}

The following real data analysis is conducted on the whole genome. Following our proposed notations we dealt with $G=2$ or two groups namely normal and cancer subjects. First, we split the genomic regions in every chromosome into windows of 100 CpG sites. We first performed Bayesian NCS fitting on the observed methylation data on 254 cancer and 32 normal patients and obtained the underlying smooth functions. In Web Appendix, supplementary figure 2 depicts the functional data visualization of the observed multivariate measurements in a genome-wide sense along all 22 chromosomes along with the combined sex chromosomes denoted as chromosome 23 from hereafter. In each plot the green and red colored curves represent the smoothed mean function for the normal and cancer patients respectively, while the blue curves denote the smoothed mean functions of all 286 patients. 

We started with window 1 by performing the smoothing spline estimation and generated 20,000 MCMC particles of $\sigma_{1,k}^2$ and $\tau_{1,k}$, $k=1,2$. For following windows, we applied our dependent method by using the transition model \eqref{lintrans} that projects the particles from window $t$ to $t+1$, for $t = 1, \dots, T-1$. As mentioned in Section \ref{sec:model} in order to account for the additional variation for the individual $j$ in group $k$ we add $\sigma_{jk}^2$, $j=1,\dots,m_k$, to the projected values of  $\widehat{\sigma}_{t+1,k}^2$ across all samples for each CpG site. We obtained Bayes factors for inference in every window as described in Section 2.3. Using a suitable threshold we determine if the data provides sufficient evidence of the presence of two distinct groups and hence detect a window to be DMR.

Figure 3 shows the DMRs detected in each window in all 23 chromosomes by our dependent FDA approach. We compare our findings with that of bumphunter approach that detects a DMR based on the number of bumps in each window. The red regions in the plots indicate the regions detected as DMRs, while the blue regions depict a non-DMR region, with darker color shades indicating extreme Bayes factor values or extreme number of bumps in a region. As can be clearly seen from the figures that the Bumphunter method  detects several DMRs in almost all the chromosomes, while our proposed dependent method, on the other hand, identifies few regions as DMR. This further validates the overestimation problem of the bumphunter method, and that the performance of our approach is quite robust. In particular, the number of DMRs detected by our approach varied from as low as 5 in chromosome 21 to as high as 81 in chromosome 1, with average being around 28. 

We further zoomed in the regions in each chromosome that had the highest concentration of DMRs and annotate the genes present in those regions. We obtained different number of genes along with gene symbols in those detected genomic locations using the information available on the \href{https://genome.ucsc.edu/cgi-bin/hgTables}{UCSC} genome browser. 

\begin{figure}[H]
\begin{center}
\includegraphics[width=18cm,height=18cm]{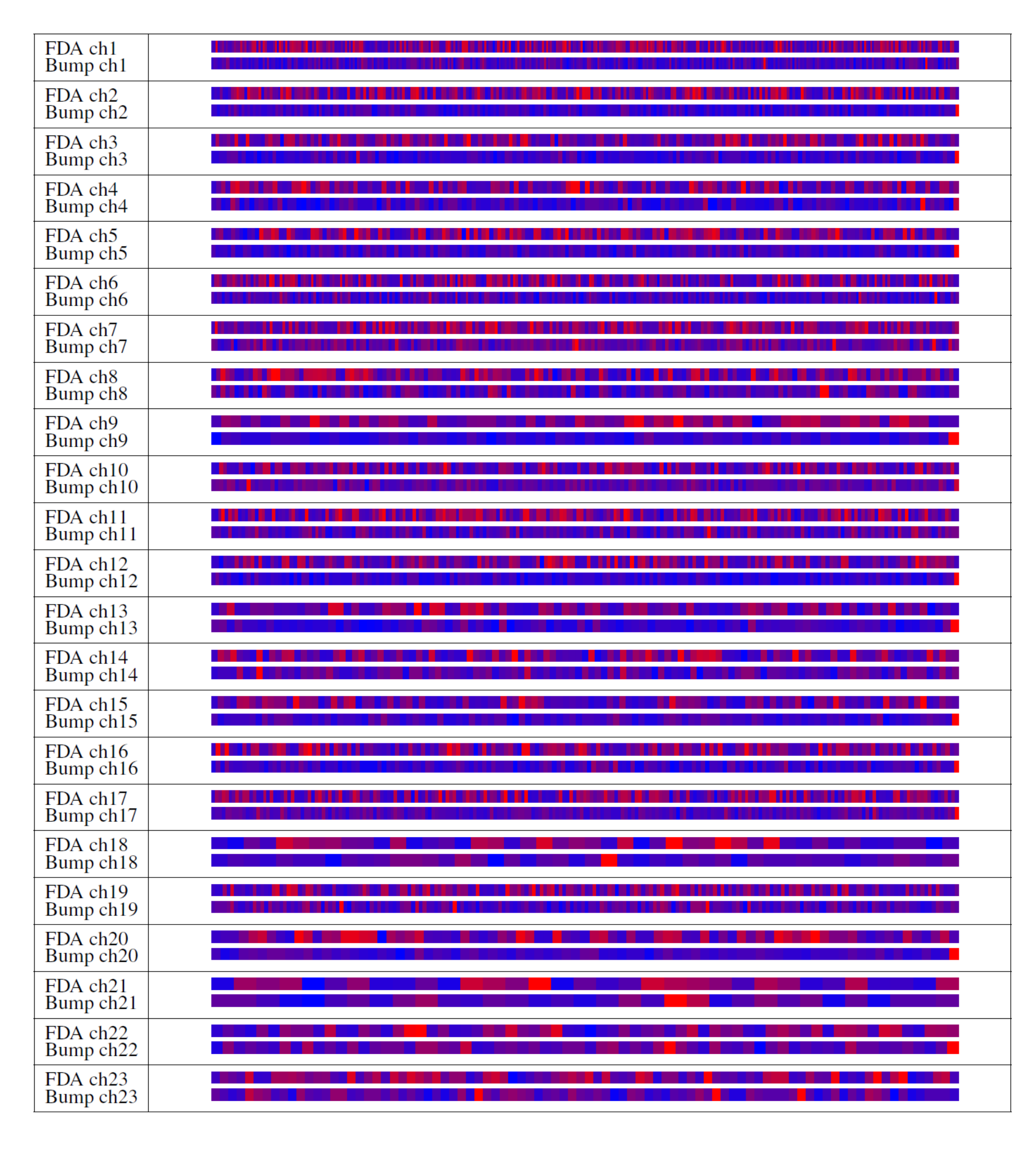}
\caption{Comparison from chromosomes 1-23 between dependent and Bumphunter method in detecting DMRs for the LUAD data. The red regions represent DMRs while blue regions are non-DMRs.} 
\label{figure3}
\end{center}
\end{figure}

In particular few genes, among all, are of interest which have been heavily cited in genomic literature and are known to be associated with lung cancer under the differential methylation in their promoter regions. Some of them include 
Epidermal  Growth  Factor Receptor that is responsible for encoding the protein which acts as a receptor for the trans-membrane glycoprotein family \citep{
81};
Kirsten RAS oncogene which is responsible for encoding the protein small GTPase superfamily \citep{
83}; 
Serine/threonine kinase 11 is responsible for encoding the protein that regulates cell polarity and is itself a tumor suppressor gene \citep{
85}; 
NeuroFibromin 1 which is a tumor suppressor gene and functions to negatively regulate the ras signal transduction pathway \citep{
84};  and 
Low-density lipo-protein Receptor-related protein 1B is responsible for encoding the protein which acts as a receptor from the low density lipo-protein
receptor family \citep{
79}. Hypermethylation of these genes are known to be associated with lung cancer, which further validates the relevance of our approach in the genome-wide identification of differentially methylated regions.

\section{Discussion}

This research is motivated from a growing body of literatures focusing on epigenetic features that may be associated with the disparities in 
Non Small Cell Lung Cancer progression and survival outcomes. The hypermethylation of the CpG island sequences located in the promoter regions of genes are increasingly being used to study the impact of epigenetic modifications. In this article we proposed Bayesian functional data analysis model to identify, select, and jointly model differential methylation features from methylation array data. Till date the applications of FDA in the genomic or public health domain remains very scarce and there is still a lot of uncovered areas in genomics that can make use of such a powerful and robust method. The proposed functional modeling approach for detecting DMR is parsimonious to address the large dimensionality of whole-genome sequencing and incorporates potential correlation among neighboring regions. We proposed a dynamically weighted particle filter with Bayesian non-parametric smoothing splines for modeling individual functional patterns followed by identification of differentially methylated regions.

We used simulation studies to compare the performance of our method with the popularly used existing bumphunter method. First we proposed our independent approach of DMR detection that fits a Bayesian NCS in individual windows without taking into account the correlation among the CpG sites from neighboring windows. As our simulation results indicated although this approach outperformed the existing bumphunter method, its performance deteriorated with increase in the correlation among the CpG sites from neighboring windows. Moreover, this approach was also challenged with high computational time. To remedy this two immediate problems associated with the independent approach we proposed our dependent approach next. In this approach we used transition models to account for the dependency that inherently exists between two genomic regions. We used an efficient sequential Monte Carlo method named dynamically weighted particle filter to get the parameter estimates of the subsequent regions without fitting the non-parametric regression function in every window. This maneuver not only made this approach computationally very efficient but also showed a very robust performance in detecting DMRs, as our simulation results indicated. This further creates a major milestone of the use of functional data modeling in genomics data.  

We applied our dependent approach to identify whole genome-wide differential methylation in Lung Adenocarcinoma cancer patients data from 
The Cancer Genome Atlas (TCGA) Program portal. We identified several DMRs along the whole genome and successfully annotated several genes in those regions, that have been reported in the literature to be associated with lung adenocarcinoma and other cancers under hyper-methylation in their promoter regions. These biological findings can further be translated into clinical research and thus we see great promises of functional data analysis in genomics data applications. 

\backmatter

\section*{Acknowledgements}
Authors are thankful to editor, associate editors and anonymous reviewers. SC was supported by the Intramural Research Program of \textit{Eunice Kennedy Shriver} National Institute of Child Health and Human Development (NICHD) of the National Institutes of Health (NIH).

\bibliographystyle{biom} 
\bibliography{ref}

\section*{Supporting Information}
Web Appendix A, referenced in Section, is available with
this paper at the Biometrics website on Wiley Online
Library.\vspace*{-8pt}

\label{lastpage}

\end{document}


\newdimen\jot \jot=5mm
\newcommand{\fnote}[1]{\mbox{$\mbox{}^{#1}$}}
\newcommand{\bigsp}{\mbox{$\; \; \; \; \;$}}
\newcommand{\half}{\mbox{$\frac{1}{2}$}}
\newcommand{\bm}[1]{ \mbox{\boldmath $ #1 $} }

\newcommand {\red}[1]{\textcolor{red}{#1}}
\newcommand {\purple}[1]{\textcolor{purple}{#1}}
\newcommand {\blue}[1]{\textcolor{blue}{#1}}
\newcommand {\green}[1]{\textcolor{green}{#1}}

\def \bx {\mbox{\boldmath $x$}}
\def \bX {\mbox{\boldmath $X$}}
\def \bY {\mbox{\boldmath $Y$}}
\def \by {\mbox{\boldmath $y$}}
\def \bbeta {\mbox{\boldmath $\beta$}}
\def \bomega {\mbox{\boldmath $\omega$}}
\def \bone {\mbox{\boldmath $1$}}
\def \L {\mbox{\boldmath $L$}} \def \G {\mbox{\boldmath $G$}}
\def \Y {\mbox{\boldmath $Y$}} \def \H {\mbox{\boldmath $H$}}
\def \W {\mbox{\boldmath $W$}} \def \M {\mbox{\boldmath $M$}}  
\def \I {\mbox{\boldmath $I$}}
\def \J {\mbox{\boldmath $J$}} \def \D {\mbox{\boldmath $D$}}
\def \K {\mbox{\boldmath $K$}} \def \Q {\mbox{\boldmath $Q$}}
\def \R {\mbox{\boldmath $R$}} \def \O {\mbox{\boldmath $O$}}
\def \e {\mbox{\boldmath $e$}}
\def \x {\mbox{\boldmath $x$}} \def \y {\mbox{\boldmath $y$}} \def \z
{\mbox{\boldmath $z$}}
\def \u {\mbox{\boldmath $u$}} \def \m {\mbox{\boldmath $m$}}
\def \F {{\cal F}}

\date{}
 \maketitle

\begin{figure}[H]
\begin{center}
\includegraphics[width=16cm,height=14cm]{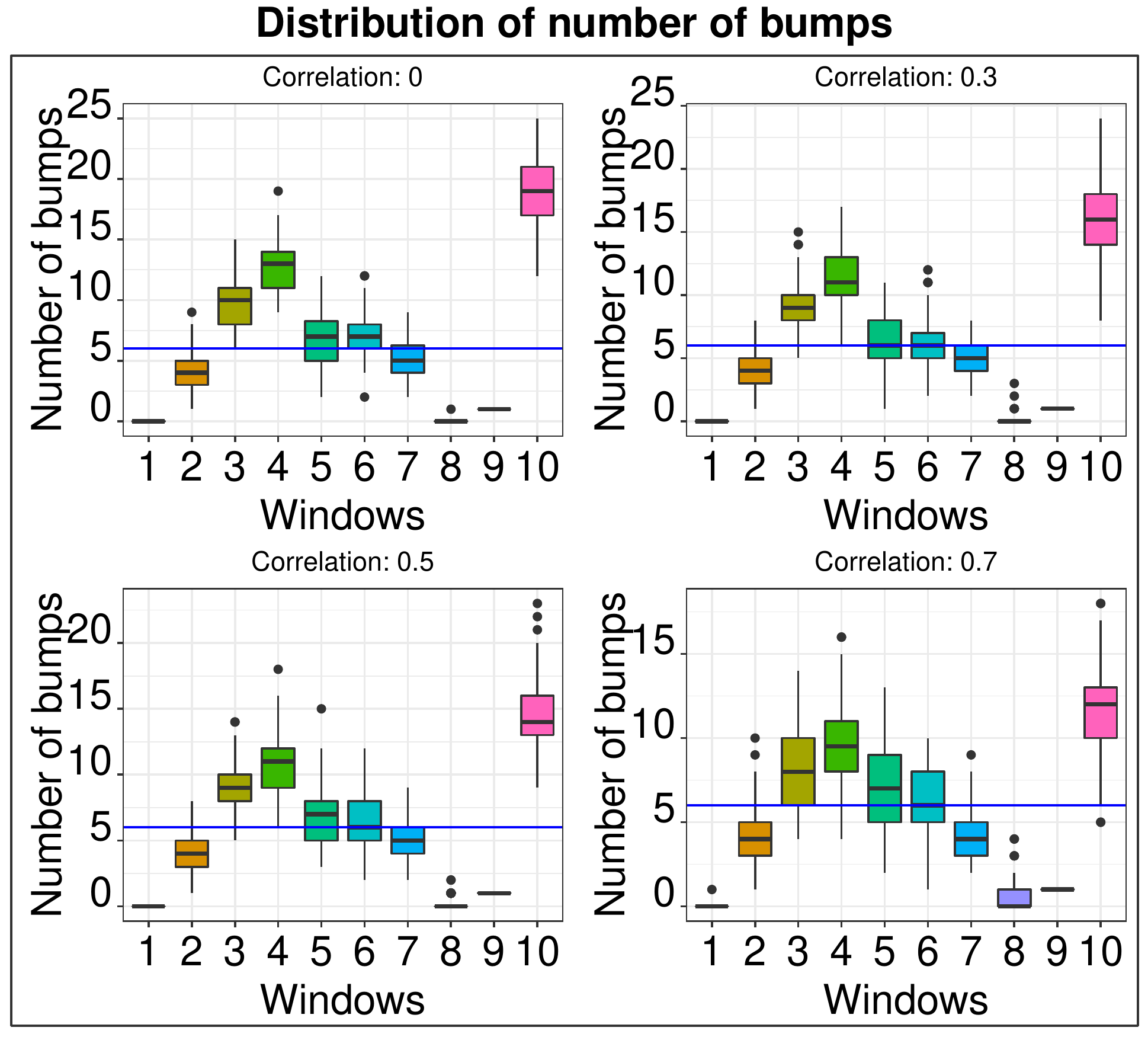}
\caption{Performance of the Bump-hunting method. The plots show the distribution of the number of bumps found for window 1 through 10 over 100 simulations in each of the 4 data sets, from top left to bottom right, with respective correlations $\rho$=0, 0.3, 0.5, 0.7. The horizontal blue line indicates the threshold value which is the average number of bumps set at 6 and any window above this line is detected as a DMR.} 
\label{figure1}
\end{center}
\end{figure}

\begin{figure}[H]
\begin{center}
\includegraphics[width=18cm,height=16cm]{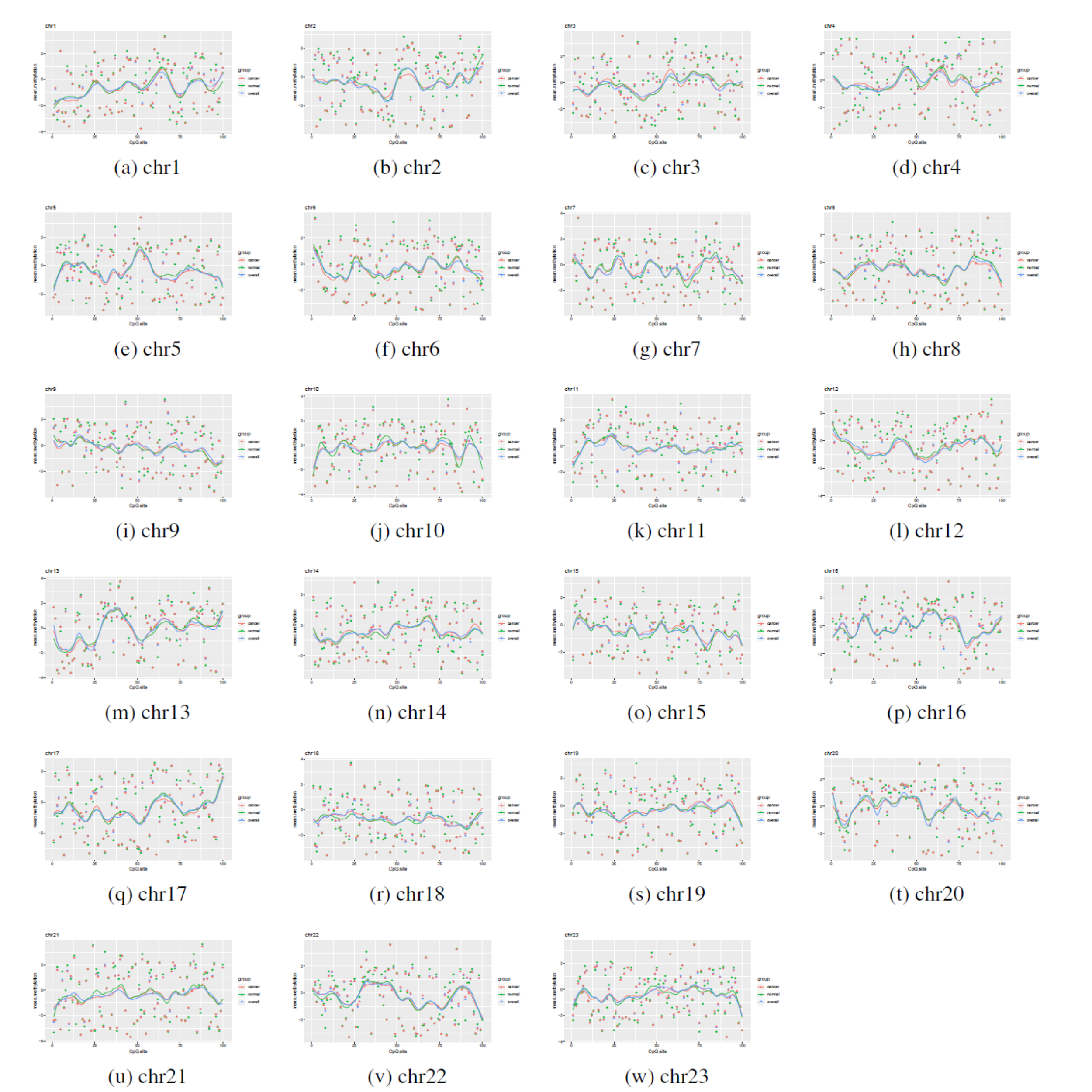}
\caption{Mean methylation rates and fitted curves for the first window in all chromosomes. Dots indicate mean methylation rates and lines indicate fitted curves. Red dots and lines are for cancer tissues, green dots and lines are for normal tissues, and blue dots and lines are for overall, respectively.} 
\label{figure2}
\end{center}
\end{figure}